\documentclass[12pt]{iopart}

\usepackage{cite}
\usepackage{enumerate}
\usepackage{amssymb}
\usepackage{amsbsy}
\usepackage[dvips]{graphicx}
\usepackage{color}
\usepackage{graphics}

\newcommand{\be}{\begin{equation}} \newcommand{\ee}{\end{equation}}
\newcommand{\bea}{\begin{eqnarray}} \newcommand{\eea}{\end{eqnarray}}
\newcommand{\el}{\nonumber \\}
\newcommand{\re}[1]{(\ref{#1})}

\newcommand{\pat}{\partial}

\renewcommand{\sec}[1]{section \ref{#1}}

\newcommand{\brt}[1]{[#1]}
\newcommand{\para}{\paragraph}

\renewcommand{\a}{\alpha}
\renewcommand{\b}{\beta}
\renewcommand{\c}{\gamma}
\renewcommand{\d}{\delta}
\renewcommand{\l}{\lambda}

\newcommand{\LCDM}{$\Lambda$CDM\ }
\newcommand{\GN}{G_{\mathrm{N}}}
\newcommand{\ha}{\frac{1}{2}}

\newcommand{\adot}{\dot{a}}
\newcommand{\addot}{\ddot{a}}
\newcommand{\rhodot}{\dot{\rho}}
\newcommand{\Hdot}{\dot{H}}

\newcommand{\bx}{\bi{x}}

\renewcommand{\H}{\frac{\adot}{a}}
\newcommand{\HH}{\frac{\adot^2}{a^2}}

\newcommand{\thetat}{\tilde{\theta}}
\newcommand{\sigmat}{\tilde{\sigma}}
\newcommand{\htt}{\tilde{h}}
\newcommand{\patl}[1]{\pat_\l#1}

\newcommand{\av}[1]{\langle{#1}\rangle}
\newcommand{\sQ}{\mathcal{Q}}
\newcommand{\sR}{{^{(3)}R}}

\newcommand{\PRD}[1]{{\it Phys. Rev.} {\bf D#1}}

\renewcommand{\PRL}[1]{{\it Phys. Rev. Lett.} {\bf #1}}

\newcommand{\PLA}[1]{{\it Phys. Lett.} {\bf A#1}}

\newcommand{\MNRAS}[1]{{\it Mon. Not. Roy. Astron. Soc.} {\bf #1}}
\newcommand{\APJ}[1]{{\it Astrophys. J.} {\bf #1}}
\newcommand{\APJS}[1]{{\it Astrophys. J. Suppl.} {\bf #1}}

\renewcommand{\CQG}[1]{{\it Class. Quant. Grav.} {\bf #1}}
\newcommand{\GRG}[1]{{\it Gen. Rel. Grav.} {\bf #1}}
\newcommand{\AaA}[1]{{\it Astron. \& Astrophys.} {\bf #1}}
\newcommand{\PROG}[1]{{\it Prog. Theor. Phys.} {\bf #1}}

\newcommand{\IJMPA}[1]{{\it Int. J. Mod. Phys.} {\bf A#1}}
\newcommand{\IJMPD}[1]{{\it Int. J. Mod. Phys.} {\bf D#1}}
\newcommand{\PRT}[1]{{\it Phys. Rept.} {\bf #1}}

\begin{document}

\begin{titlepage}

\title{Light propagation in statistically homogeneous and isotropic dust universes}

\author{Syksy R\"{a}s\"{a}nen}

\address{Universit\'e de Gen\`eve, D\'epartement de Physique Th\'eorique \\
24 quai Ernest-Ansermet, CH-1211 Gen\`eve 4, Switzerland}

\ead{syksy {\it dot} rasanen {\it at} iki {\it dot} fi}

\begin{abstract}

\noindent We derive
the redshift and the angular diameter
distance in rotationless dust universes
which are statistically homogeneous and isotropic,
but have otherwise arbitrary geometry.
The calculation from first principles
shows that the Dyer-Roeder approximation does
not correctly describe the effect of clumping.
Instead, the redshift and the distance are
determined by the average expansion
rate, the matter density today and the null geodesic shear.
In particular, the position of the CMB peaks
is consistent with significant spatial curvature
provided the expansion history is sufficiently
close to the spatially flat \LCDM model.

\end{abstract}


\end{titlepage}

\setcounter{tocdepth}{2}

\tableofcontents

\setcounter{secnumdepth}{3}

\section{Introduction} \label{sec:intro}

\para{Observations and clumpiness.}

Observations of the universe
are inconsistent with homogeneous and isotropic
cosmological models with ordinary matter and standard gravity
(meaning matter with non-negative pressure and
gravity that is described by the four-dimensional Einstein-Hilbert action).
Observational results are usually expressed in terms of their interpretation
in the context of the homogeneous and isotropic Friedmann-Robertson-Walker
(FRW) models, in which there is a one-to-one relationship between the
expansion rate and the distance (given the spatial curvature).
Analysed this way, observations imply that the expansion has
accelerated in the past few billion years \cite{FRWexp, Cattoen}.
A model-independent statement would be that the observed distances
at late times are a factor of about 2 higher than expected in FRW
models with ordinary matter and gravity.
This is usually taken as evidence for exotic matter
with negative pressure or a modification of gravity.
In particular, the homogeneous and isotropic, spatially flat
model \LCDM model fits observations of the distance scale
and the expansion rate well by introducing vacuum energy.
(We will refer to all homogeneous and isotropic models which
contain only dust and vacuum energy as $\Lambda$CDM, whatever
the spatial curvature.)
However, the universe is known to be far from exact homogeneity
and isotropy at late times due to the formation of non-linear
structures. Before concluding that new physics is needed, it is
necessary to quantify the effect of clumpiness on
the observations.

The influence of inhomogeneity and/or anisotropy on the average
evolution was first mentioned in \cite{Shirokov:1963} and was
discussed in detail in \cite{fitting} under the name
``fitting problem''. The effect on the
expansion rate is also known as backreaction
\cite{Buchert:1995, Buchert:1999, Buchert:2001}; see
\cite{Ellis:2005, Rasanen:2006b, Buchert:2007} for reviews.
The possibility that backreaction could lead to accelerated
expansion and explain the observations without new
physics was first considered in \cite{Wetterich:2001, Schwarz}
(and briefly mentioned in \cite{Buchert:2000, Tatekawa:2001}).
Both metric and matter perturbations were taken into account
and the observables were correctly identified in \cite{Rasanen},
where first order perturbation theory was expanded to second
order, and this was extended to a
consistent second order calculation in \cite{Kolb:2004a}.
In \cite{Kai:2006, Rasanen:2006a, Rasanen:2006b} it was explained
with toy models that the physical reason for
average acceleration is that faster expanding
regions come to occupy a larger fraction of the volume.
Accelerated expansion has also been explicitly demonstrated
with the exact Lema\^{\i}tre-Tolman-Bondi (LTB) solution
\cite{Kai:2006, Chuang:2005, Paranjape:2006a}.
A semi-realistic model with an evolving distribution of non-linear
structures was studied in \cite{Rasanen:2008a, peakrevs},
and it was found that the expansion rate rises (relative to
the homogeneous and isotropic case) by the right order of
magnitude around the right time, some billions to tens of
billions of years, though not rapidly enough to correspond
to acceleration. The model involved several approximations,
and a more careful treatment would be needed for detailed
comparison with observations.

However, most observations, including those of the cosmic
microwave background (CMB) \cite{Komatsu:2008}
and type Ia supernovae \cite{SNobs}, do not
directly measure the expansion rate, but rather
cosmological distances and redshifts, which are
defined in terms of light propagation.
The few measurements that are sensitive to the expansion rate
independent of the distance scale are those of the
local Hubble parameter \cite{Hubble, Jackson:2007},
the ages of passively evolving galaxies as a function
of redshift \cite{ages}, the Integrated Sachs-Wolfe (ISW)
effect \cite{Ho:2008, Giannantonio:2008}
and the growth rate of matter fluctuations \cite{lineargrowth}.
Baryon acoustic oscillations depend on a mixture of the
expansion rate and distance \cite{BAO1, BAO2}.

\para{The distance and the expansion rate.}

In addition to changing the expansion of the universe, non-linear
structures affect the relationship between the expansion rate
and light propagation.
In a general spacetime, there is no direct relationship between
the expansion rate and the distance scale, and it would in
principle be possible to explain the observations without accelerated
expansion. For example, in models where we are located near the
center of an untypically large spherical void, the distances
can be consistent with the observations, but generally
there is no acceleration \cite{Enqvist:2006, Bolejko:2008c}.
(See \cite{Enqvist:2007} for a review,
\cite{Rasanen:2006b} for more references, and
\cite{Caldwell:2007, Uzan:2008, GarciaBellido:2008a, Bolejko:2008b, Zibin:2008b, GarciaBellido:2008b}
for observational constraints related to the inhomogeneity of these models.)
Even if the large local void models are not realistic, studying
them with the exact LTB solution has established unambiguously
that inhomogeneities with density contrast of order unity
and sizes smaller than the horizon can have a large
impact on the distance scale. This is in contrast to
perturbation theory arguments based on the amplitude
of metric perturbations in the longitudinal gauge.

The speculative possibility of an unexpectedly large local void
aside, the observed universe seems to be statistically
homogeneous and isotropic on large scales, with a homogeneity
scale of around 100 Mpc \cite{Hogg:2004, Pietronero}
(though see \cite{morphology, SylosLabini}).
(For discussion of statistically homogeneous and isotropic
but locally clumpy dust universes, see \cite{Rasanen:2006b, Rasanen:2008a}.)
The relevant question is then what is the effect of such
a distribution of non-linear structures on light propagation.

It was conjectured in \cite{Rasanen:2008a}
that in a statistically homogeneous and isotropic dust universe,
light propagation can be treated in terms of the overall geometry
(meaning the average expansion rate and average spatial curvature)
if the structures are realistically small
and the observer is not in a special location.
Such a conjecture is in agreement with various studies of
light propagation (see \cite{Rasanen:2008a} for an
overview and references), and it is also suggested
by the fact that different observations are well explained
in terms of the evolution of a single scale factor.
However, until now there had been no proof of the conjecture, and
it was not known whether additional conditions are necessary
in addition to statistical homogeneity and isotropy.

In the present work, we clarify the relationship between the
expansion rate and the distance scale in statistically homogeneous
and isotropic dust universes which may contain non-linear structures.
We relate the redshift and the distance to the dust geometry,
and confirm that light propagation can be expressed in terms of
average geometrical quantities, up to a term related to the null
geodesic shear.
In fact, the null shear aside, the average expansion rate
(and the matter density today)
is sufficient to determine the distance, and the spatial
curvature enters only via its effect on the expansion rate.
This implies that a clumpy model can be consistent with the
observed position of the CMB acoustic peaks even when there
is significant spatial curvature.
The result also shows that the Dyer-Roeder prescription of multiplying
the matter density by a smoothness factor does not
correctly describe the effect of clumping.

In section 2 we set up the dust geometry, in section 3 we
relate the redshift and the distance to the average geometry,
and in section 4 we discuss and summarise our results.

\section{The dust geometry}

\subsection{The local equations}

\para{The gradient decomposition.}

We are interested in light propagation in a dust spacetime.
The geometry is a solution to the Einstein equation with dust matter,
\bea \label{Einstein}
  R_{\a\b} - \frac{1}{2} g_{\a\b} R &=& 8 \pi\GN T_{\a\b} \el
  &=& 8 \pi\GN \rho\, u_{\a} u_{\b} \ ,
\eea

\noindent where $R_{\a\b}$ is the Ricci tensor,
$R$ is the Ricci scalar, $\GN$ is Newton's constant,
$T_{\a\b}$ is the energy-momentum tensor,
$\rho$ is the dust energy density and $u^{\a}$ is the velocity
of observers comoving with the dust.

The velocity has unit norm, $u_\a u^\a=-1$.
Since the pressure is zero, $u^\a$ is the tangent vector of
timelike geodesics, $u^\b\nabla_\b u^\a=0$.
We can define a tensor which  projects on the tangent space
orthogonal to $u^\a$ by
\bea \label{threemetric}
  h_{\a\b} \equiv g_{\a\b} + u_\a u_\b \ ,
\eea

\noindent where $g_{\a\b}$ is the full metric. The
three-metric $h_{\a\b}$ satisfies $h_{\a\b}u^\b=0$, 
$h_{\a}^{\ \c} h_{\c\b}=h_{\a\b}$, $h^{\a}_{\ \a}=3$.

It is useful to express the geometry in terms of the
decomposition of the covariant derivative of $u^\a$
(for reviews of this covariant approach, see
\cite{Ehlers:1961, Ellis:1971, Wainwright:1997, Ellis:1998c, Tsagas:2007}):
\bea \label{dustdec}
  \nabla_\b u_\a &=& \theta_{\a\b} + \omega_{\a\b} \el
  &=& \frac{1}{3} h_{\a\b} \theta + \sigma_{\a\b} + \omega_{\a\b} \ ,
\eea

\noindent where the symmetric part $\theta_{\a\b}=\nabla_{(\b} u_{\a)}$
is the expansion tensor and the antisymmetric part
$\omega_{\a\b} = \nabla_{[\b} u_{\a]}$ is the vorticity tensor.
The trace of the expansion tensor \mbox{$\theta=\nabla_\a u^\a$}
is the volume expansion rate and the trace-free part
$\sigma_{\a\b} = \nabla_{(\a} u_{\b)}-\frac{1}{3} h_{\a\b} \theta$
is the shear tensor.
Like $h_{\a\b}$, the tensors $\sigma_{\a\b}$ and $\omega_{\a\b}$
are spatial in the sense that they are orthogonal to $u^\a$,
\mbox{$\omega_{\a\b}u^\b=\sigma_{\a\b}u^\b=0$}.
They are also traceless, \mbox{$\omega^{\a}_{\ \a}=\sigma^{\a}_{\ \a}=0$}.

\para{The scalar equations.}

The Einstein equation \re{Einstein} can be conveniently written in
terms of the decomposition \re{dustdec}
and the electric and magnetic parts of the Weyl tensor. 
For the full system of equations, see \cite{Wainwright:1997} (page 27).
We are interested in the overall geometry, in other words in
average quantities. Since only scalars can be straightforwardly
averaged in a curved spacetime (though see
\cite{Carfora, Zalaletdinov, Paranjape}), we will consider
only the scalar part of the Einstein equation, which reads
\bea
  \label{Rayloc} \dot{\theta} + \frac{1}{3} \theta^2 &=& - 4 \pi \GN \rho - 2 \sigma^2 + 2 \omega^2 \\
  \label{Hamloc} \frac{1}{3} \theta^2 &=& 8 \pi \GN \rho - \frac{1}{2} \sR + \sigma^2 - \omega^2 \\
  \label{consloc} \rhodot + \theta\rho &=& 0 \ ,
\eea

\noindent where a dot stands for $\pat_t\equiv u^\a\nabla_\a$,
the covariant derivative with respect to proper time $t$ measured
by observers comoving with the dust,
$\sigma^2=\ha\sigma_{\a\b}\sigma^{\a\b}\geq0$ is the shear scalar,
$\omega^2=\ha\omega_{\a\b}\omega^{\a\b}\geq0$ is the vorticity scalar,
and $\sR$ is the Ricci scalar on the tangent space orthogonal to $u^\a$.
The acceleration equation \re{Rayloc} is known as the Raychaudhuri
equation, and \re{Hamloc} is the Hamiltonian constraint.

\subsection{The average equations} \label{sec:av}

\para{Defining the average.}

If and only if the vorticity is zero, the tangent spaces
orthogonal to $u^\a$ form spatial hypersurfaces, and provide
a foliation that fills the spacetime exactly once. These
flow-orthogonal hypersurfaces coincide with the hypersurfaces
of constant proper time of comoving observers. If the vorticity
is non-zero, the hypersurfaces of constant proper time are no
longer orthogonal to the fluid flow \cite{Ehlers:1961, Ellis:1971}.

We assume in this subsection that the vorticity is zero, and
follow the formalism of \cite{Buchert:1999, Buchert:2001}.
The spatial average of a quantity is then its integral over the
hypersurface of constant proper time $t$ orthogonal to $u^\a$,
divided by the volume of the hypersurface,
\bea \label{av}
  \av{f}(t) \equiv \frac{ \int_t \epsilon f }{ \int_t \epsilon } \ ,
\eea

\noindent where $\epsilon_{\a\b\c}\equiv\eta_{\a\b\c\d} u^\d$
is the volume element on the tangent space orthogonal
to $u^\a$, $\eta_{\a\b\c\d}$ being the spacetime volume element.

In particular, the average expansion rate is
\bea \label{avtheta}
  \av\theta(t) &=& \frac{ \int_t \epsilon\, \theta }{ \int_t \epsilon } \el
  &=& \frac{ \pat_t \int_t \epsilon }{ \int_t \epsilon } \el
  &\equiv& 3 \frac{\adot}{a} \ ,
\eea

\noindent where we have defined the scale factor $a(t)$ as the volume
of the hypersurface of constant proper time to power $1/3$,
\bea \label{a}
  a(t) \equiv \left( \frac{ \int_t \epsilon}{ \int_{t_0} \epsilon} \right)^{\frac{1}{3}}  \ ,
\eea

\noindent and $a$ has been normalised to unity at time $t_0$,
which we take to be today. We will also use the notation
$H\equiv\adot/a$.

\paragraph{The Buchert equations.}

Let us take the average of the equations \re{Rayloc}--\re{consloc}.
The resulting Buchert equations are \cite{Buchert:1999}
\bea
  \label{Ray} 3 \frac{\addot}{a} &=& - 4 \pi\GN \av{\rho} + \sQ \\
  \label{Ham} 3 \HH &=& 8 \pi \GN \av{\rho} - \frac{1}{2}\av{\sR} - \frac{1}{2}\sQ \\
  \label{cons} && \pat_t \av{\rho} + 3 \H \av{\rho} = 0 \ ,
\eea

\noindent where the backreaction variable $\sQ$ contains the effect
of inhomogeneity and anisotropy:
\bea \label{Q}
  \sQ \equiv \frac{2}{3}\left( \av{\theta^2} - \av{\theta}^2 \right) - 2 \av{\sigma^2} \ .
\eea

\noindent The integrability condition between \re{Ray} and \re{Ham}
reads
\bea \label{int}
  \pat_t {\av{\sR}} + 2 \H \av{\sR} = - \pat_t \sQ - 6 \H \sQ \ .
\eea

The Buchert equations \re{Ray}--\re{cons} describe the evolution
of the volume $a^3$ of a spatial domain, or equivalently
its average expansion rate $\av\theta$.
They differ from the FRW equations by the presence of the
backreaction variable $\sQ$, and the related fact that the
average spatial curvature $\av\sR$ can evolve in a non-trivial
manner, as indicated by the integrability condition \re{int},
whereas in FRW universes it is always proportional to $a^{-2}$
\cite{Misner:1973} (page 720), \cite{Rasanen:2007}.
If the backreaction variable $\sQ$ is large enough, the expansion
will accelerate, as indicated by \re{Ray}, and the spatial curvature
will be correspondingly large, as indicated by \re{Ham} and \re{int}
\cite{Rasanen:2005, Rasanen:2006b}.

\section{The light propagation}

\subsection{The redshift} \label{sec:redshift}

\para{Geometrical optics.}

When the wavelength of light is much shorter than both the
local curvature radius and the typical scale over which the
amplitude and the wavelength change appreciably, light
propagation can be treated in the geometrical optics approximation
\cite{Misner:1973} (page 570), \cite{Schneider:1992} (page 93).
In geometrical optics, light travels along null geodesics,
and the light rays have no effect on the geometry.
The tangent vector of the null geodesics is given by the
gradient of the wave phase $S$, and it is identified with the photon
momentum, $k_\a=\pat_\a S$. 
The null geodesic tangent vector satisfies
$k_\a k^\a=0$ and $k^\a \nabla_\a k^\b=0$.

We will consider the propagation of a bundle of nearby null
geodesics. We are interested in the redshift and the surface
area of the bundle, the latter of which gives the angular
diameter distance.
We will not consider caustics, which are not expected
to be important for typical light rays in cosmology
(though see \cite{caustic}).

\para{The general expression for the redshift.}

The spacetime geometry is determined dynamically by the
Einstein equation \re{Einstein} and traced by the dust
geodesics with tangent vector $u^\a$.
Light propagation involves a derived geometrical structure,
given by the solution of the null geodesic equations in
the fixed spacetime geometry, traced by the photon
geodesics with tangent vector $k^\a$.
The tangent vectors $u^\a$ and $k^\a$ are parallel propagated
with respect to the dust and photon geodesics, respectively,
but not with respect to each other. It follows that
the photon momentum changes along the dust geodesics.
The redshift $z$ of a source is defined as the observed
photon wavelength divided by the wavelength at the source, minus
one. The wavelength is inversely proportional to the energy $E$, so
\bea \label{z}
  1 + z &=& \frac{E_{\mathrm{s}}}{E_{\mathrm{o}}} \ ,
\eea 

\noindent where s refers to the source and o to the observer.
The energy is the projection of the momentum onto the observer's
velocity, given by the tangent vector of the dust geodesic,
\bea \label{E}
  E = - u_\a k^\a \ .
\eea

It is convenient to decompose $k^\a$
into an amplitude and the direction, and split the direction into
components orthogonal and parallel to the dust geodesics,
\bea \label{kdec}
  k^\a = E ( u^\a + e^\a ) \ ,
\eea

\noindent where $u_\a e^\a=0$, $e_\a e^\a =1$. The vector $e^\a$
is spatial in the sense that it lies in the three-space
which has the metric $h_{\a\b}$, $h^{\a}_{\ \b} e^\b=e^\a$.

To find out how the energy evolves along the null geodesic,
we take the derivative with respect to the affine parameter $\lambda$,
\bea \label{Eeq}
  \patl{E} &\equiv& k^\a \nabla_\a E \el
  &=& - k^\a k^\b \nabla_\a u_\b \el
  &=& - k^\a k^\b ( \theta_{\b\a} + \omega_{\b\a} ) \el
  &=& - E^2 e^\a e^\b \theta_{\a\b} \el
  &=& - E^2 \left( \frac{1}{3} \theta + \sigma_{\a\b} e^\a e^\b \right) \ ,
\eea

\noindent where we have applied the decompositions \re{dustdec}
and \re{kdec} and taken into account $k^\a \nabla_\a k^\b=0$.
Note that the vorticity tensor drops out due to its asymmetry.
We can integrate \re{Eeq} to obtain
$E\propto \exp\left( - \int\rmd\l E \, \theta_{\a\b} e^\a e^\b \right)$
(the factor $E$ is retained in the integrand for later convenience).
Using \re{z} we then have for the redshift
\bea \label{zsolution}
  1 + z &=& \exp\left( \int_{\l_\mathrm{s}}^{\l_\mathrm{o}} \rmd\l E \theta_{\a\b} e^\a e^\b \right) \el
  &=& \exp\left( \int_{\l_\mathrm{s}}^{\l_\mathrm{o}} \rmd\l E \left[ \frac{1}{3}\theta + \sigma_{\a\b} e^\a e^\b \right]\right) \ ,
\eea

\noindent where the integral is from the source to the observer
along a specific geodesic.
The above relation gives the redshift in a general dust spacetime
in terms of the dust geometry and the spatial direction $e^\a$ of
the null geodesics. (Vorticity enters indirectly via the geodesic
equation which determines $e^\a$.)

Looking at the redshift from the viewpoint of observers
on dust geodesics, $u^\a$ is constant, and the momentum
$k^\a$ decreases. However, following the null geodesics instead
makes the relation to the change in the dust geometry more transparent.
Along a null geodesic the momentum $k^\a$ is constant, and the
product $-u_\a k^\a$ changes because $u_\a$ is turning along the
photon path, which is precisely what $\nabla_\b u_\a$ quantifies.
(Following the dust geodesics by taking the derivative
$\pat_t=u^\a \nabla_\a$ instead of $\patl{}$ would
lead to an expression for the redshift in terms of the turning
of the null geodesics.)

Let us assume that the vorticity vanishes.
Then the hypersurfaces of constant proper time are orthogonal to $u^\a$,
and $t(\l)$ is monotonic. We can therefore parametrise
points along the geodesic with the value of the proper time $t$ on the
hypersurface that the null geodesic is crossing\footnote{When
rotation is present, it is not obvious that a light ray could not
pass from a higher value of $t$ to a smaller value.
Note that light propagating from an observer with a
larger proper time to one with a smaller proper time does not
necessarily violate causality.}.
We can then invert $\patl=E(u^\a+e^\a)\pat_\a$ to obtain
$\int\rmd\l=\int_{\l}\rmd t E^{-1}$, so we have
\bea \label{zt}
  \!\!\!\!\!\!\!\!\!\!\!\! \!\!\!\!\!\! 1 + z &=& \exp\left( \int_{t, \l}^{t_0} \rmd t \left[ \frac{1}{3}\theta(t, \bx(t)) + \sigma_{\a\b}(t, \bx(t)) e^\a(t, \bx(t)) e^\b(t, \bx(t)) \right]\right) \ ,
\eea

\noindent where the subscript $\l$ indicates that the integral
is from the source to the observer along a
\linebreak
specific geodesic, which crosses the hypersurface of proper
time $t$ at spatial position $\bx(t)$.

\para{Statistical homogeneity and isotropy.}

In a FRW universe, the shear is zero, and only the expansion rate
remains in the integral \re{zt}. Since the universe is homogeneous
and isotropic, it does not matter in which direction the geodesic
goes, and the result $1+z=a(t)^{-1}$ immediately follows from
the definition of the scale factor \re{a}.

In a general dust universe, the shear can be important, and
the result depends on the direction of the null geodesic.
In \cite{Schneider:1992} (page 136), it was assumed that
the average of $\theta_{\a\b} e^\a e^\b$ along the ray reduces to
the FRW expansion rate, and that this may be considered as
part of the incomplete definition of a model being ``on average FRW''.
We will try to make this reasoning somewhat more explicit,
and take into account that the average expansion rate
does not necessarily reduce to the FRW one.

In a general dust spacetime, there is no direct relationship
between the redshift \re{zt} and the scale factor \re{a}.
The redshift is given by the integral of the local expansion rate
and projected shear along a null geodesic, while the scale factor
is given by the time integral of the spatially averaged
expansion rate.
However, in a statistically homogeneous and isotropic
universe, the two quantities are closely related.
If structures are identical in all
directions up to statistical fluctuations, and if the coherence
scale of the distribution (related to the homogeneity scale) is much
smaller than the distance over which we consider light
propagation, then the redshift should be the same in all directions,
and it should not depend on the specific geodesic we are looking at.
We can view this in the following manner.

The dust shear $\sigma_{\a\b}$ is related to the structures through
which the light travels. If the structures have no preferred
orientation, the shear is uncorrelated with the direction of
the null geodesic, and the integral of the shear projected on the
geodesic should vanish over long distances.
In other words, the integral of $\theta_{\a\b} e^\a e^\b$
should be the same for all $e^\a$, which implies that
only the trace contributes.
Furthermore, the integral of the local expansion rate along
the null geodesic can be related to the integral of the average
expansion rate over time. Neglecting the shear,
we split the expansion rate at each point along the geodesic
into the average value on the hypersurface of constant proper time
at that point (given by \re{avtheta}) and the variation,
\bea \label{zHI}
  1 + z &=& \exp\left( \int_{t, \l}^{t_0} \rmd t \left[\frac{1}{3} \av\theta(t) + \frac{1}{3} \Delta\theta(t,\bx(t)) \right] \right) \ ,
\eea

\noindent where $\Delta\theta\equiv\theta(t,\bx(t))-\av\theta(t)$.
(In what follows, we will use the notation $f=\av{f}+\Delta f$ for any
scalar quantity $f$; there is no assumption that $\Delta f$ is small,
and hence no loss of generality.)
If the distribution of structures evolves slowly compared to the time
it takes for light to cross the homogeneity scale (i.e. for the null
geodesic to integrate over a statistically homogeneous and isotropic
sample), then the contribution of the variation $\Delta\theta$ should
be small compared to the contribution of the average $\av{\theta}$.
Consider two geodesics passing from the hypersurface with
proper time $t$ to one with proper time $t+\Delta t$. If the
geodesics cross the hypersurfaces at different points, they
will go through different structures on the way and will
in general gain different amounts of redshift.
However, if the structures evolve slowly
compared to the passage time $\Delta t$, the distribution of
structures is essentially static between $t$ and $t+\Delta t$.
If the distance $\Delta t$ is at least as large
as the homogeneity scale, the redshift gained is the same for
both geodesics, up to statistical fluctuations.

In the real universe, the size of structures and the homogeneity
scale are indeed small compared to the Hubble time, which is the
timescale of the evolution of the distribution of structures.
For typical supersymmetric weakly interacting dark matter
candidates, the size of the first structures, which form around 
$z\sim$ 40--60, is of the order $10^{-8} H^{-1}$ \cite{SUSYCDM}.
The size of structures relative to the Hubble length grows
as structure formation proceeds, and today the largest
typical structures have sizes of around 10 Mpc
$\approx 10^{-3} H^{-1}$.
The homogeneity scale today is of the order of
100 Mpc $\approx 10^{-2} H^{-1}$ \cite{Hogg:2004, Pietronero}
(though see \cite{morphology, SylosLabini}).

If the contribution of $\Delta\theta$ in \re{zHI} can be neglected,
the redshift does not depend on the specific geodesic, and we have
\bea \label{za}
  1 + z &\approx& 1 + \av{z} = a(t)^{-1} \ ,
\eea

\noindent where $a(t)$ is the scale factor defined in \re{a}. The average
redshift on the hypersurface of proper time $\av{z}$ can
by statistical homogeneity and isotropy also be understood as
the average along a specific geodesic taken over a distance
longer than the homogeneity scale (but much shorter than the Hubble scale).
In a statistically homogeneous and isotropic universe,
the redshift is independent of direction, and it is related to the
volume expansion rate in the same way as in the exactly homogeneous
and isotropic FRW models. The change in the wavelength of a typical
photon is, over long distances, only determined by the overall
expansion of the volume. (This conclusion disagrees with
\cite{Larena:2008, Rosenthal:2008}, where the redshift was
considered using ad hoc treatments for the spatial
curvature; see also \cite{Rasanen:2007}.)

While the relation \re{za} between the expansion rate and the
redshift is simple, the result is not entirely obvious.
For example, it is vital that the shear and the expansion rate
appear linearly in the redshift integral \re{zt}.
In comparison, the shear and the expansion rate contribute
quadratically to the dust equations of motion \re{Rayloc}
and \re{Hamloc}, so the shear does not drop out when considering
large regions, and the variance of the expansion rate
plays an important role.

The parametrisation of the null geodesics in terms of
the proper time is crucial. The evolution of structures
is governed by proper time, so the hypersurface of
proper time is also the hypersurface of statistical homogeneity and
isotropy. It is the statistical homogeneity and isotropy which
makes it possible to neglect the shear and $\Delta\theta$, and
relate the scale factor \re{a} to observables.
This expresses in more detail the relation between the redshift,
the scale factor and statistical homogeneity and isotropy
discussed in \cite{Rasanen:2006b, Rasanen:2008a}.
For making this argument sharper, it would be worthwhile to have
a more precise notion of statistical homogeneity and isotropy
in a general dust spacetime.

This connection with the observable redshift establishes
the observational relevance of the expansion rate averaged
over the hypersurface of constant proper time.
Studying the average expansion rate has been criticised
\cite{Ishibashi:2005} on the grounds that it depends on
the hypersurface of averaging \cite{Geshnizjani, Rasanen:2004},
and also because the average is taken on a spacelike hypersurface,
while observations are made along and inside the past light cone.
Though there is a preferred foliation for dust, given by the
hypersurface of constant proper time of comoving observers
\cite{Kolb:2005c, Rasanen:2006b, Rasanen:2008a},
ultimately the usefulness of the expansion rate averaged on
that hypersurface is determined by relating it to observed
quantities, which we have now done for the redshift.

The expansion rate and the shear can have large variations
along the null geodesic, of the order of the average
values. A crucial reason for why it is sufficient to consider
only the averages is that the coherence length of the
variations is much smaller than the scale over which the
averages change significantly.
(The large amplitude of local variations in the expansion
rate is clear from \re{Rayloc}--\re{consloc} and the fact
that there are large differences in the local density.
This can be seen explicitly in the spherical
collapse model and its underdense equivalent
\cite{Gunn:1972, Sheth:2003}.)

\para{Deviations from the mean.}

In a universe which is only statistically and not exactly
homogeneous and isotropic, the contribution of the shear to
the redshift integral \re{zt} is not zero, only suppressed
compared to the contribution of the expansion rate.
The fact that the null geodesics are affected by the dust shear
also leads to a correlation between the shear and the spatial direction
of the null geodesic $e^\a$.
However, from observations it is known that the change in
$e^\a$ for typical light rays is small, at the percent level
\cite{Munshi:2006}.
Similarly, because there are statistical fluctuations and
because the distribution is not exactly static, the integral
of the spatial variation of the expansion rate will not
completely vanish even over scales much longer than the
homogeneity scale, though its contribution to \re{zHI}
will be small compared to the contribution of the average
expansion rate.
In perturbed FRW models, the contribution of the shear
and $\Delta\theta$ reduces, in addition
to the local dipole term, to the ISW effect and the Rees-Sciama effect.
(For covariant treatment of CMB perturbation theory, see \cite{Dunsby:1997}.
The covariant formalism was also recently applied to the CMB in
\cite{Zibin:2008a}.)

Another source of variation is the fact that the cancellations
discussed above only happen when integrating over long
distances. For distances smaller than the homogeneity scale,
directional variation in the redshift due to differences in
the local expansion rate and shear should be expected.
Typical variation of the expansion rate in different directions
within 70 Mpc around our location was found to be 20\% in
\cite{McClure:2007} (see also \cite{dirvar}).
For the nearby supernovae, these variations are well known
under the name peculiar velocities. (In the comoving approach we follow,
peculiar velocities are zero by definition. For a covariant way of
defining peculiar velocities, see \cite{peculiar}.)
Because the homogeneity scale is small compared to the distance
over which most cosmological observations are made, the effect
of these deviations is expected to be small.

A rough estimate of the contribution of the local variation
in the geometry to the redshift \re{zt} would be
$\int \rmd t \Delta\theta\approx L \av{\Delta\theta}$,
where $L$ is the size of the region where there is significant
uncancelled variation, and $\av{\Delta\theta}$ is the typical
magnitude of the variation. For $L=70$ Mpc and
$|\av{\Delta\theta}|= 0.2 H_0$, where $H_0$ is the Hubble
parameter today, we get $10^{-3}$.
This may be an overestimate, because in the linear
regime perturbations in the expansion rate and the
shear mostly cancel, apart from the dipole
(for which the contribution of local structures is
indeed observed to be $10^{-3}$).
There does not appear to be a reason for such a cancellation
in the non-linear regime, but the importance of the deviation from
the linear results for the local universe remains to be
determined\footnote{In the average expansion rate, the deviation
from FRW behaviour does not become significant until fourth order,
because of a similar cancellation between the local expansion
rate and the shear at second order. This can be understood
in terms of the Newtonian limit \cite{Notari:2005, Rasanen:2006b}. 
It is not clear whether the cancellation in the redshift can
be understood in a similar manner.}.
In any case, the correction is negligible for cosmological probes other
than the CMB. For the CMB, the effect might give an important
contribution at large angles (where the variation in the expansion
rate is large), and could possibly explain some of the observed
anomalies \cite{Huterer:2006, Copi:2008} (see
\cite{Rasanen:2006b} for discussion and more references),
particularly since some of them are correlated with the dipole.

Observationally, we know from the CMB that the redshift of
the last scattering surface is the same in all directions,
i.e. for different geodesics, to the level $10^{-3}$,
or $10^{-5}$ if we exclude the dipole.
Assuming that our location is typical in space, this
is the variation over the hypersurface of constant proper time.
The variation is likely be at least as small at earlier
times, because structure formation is less advanced.
In addition, deviation of the CMB spectrum from the
blackbody shape is observed to be small \cite{Fixsen:1996}.
The sum of two blackbody spectra at different temperatures
is not a blackbody, so finite angular resolution inevitably
changes the spectral shape when different directions have
different CMB temperatures \cite{Chluba:2004}.
Scattering of the CMB leads to a similar effect \cite{Stebbins:2007}
(this was used to constrain local
void models in \cite{Caldwell:2007}).
The limit on the $y$-distortion relevant for both
cases is $|y|<1.5\times 10^{-5}$ \cite{Fixsen:1996}.
So the cancellations discussed above are well realised
in the universe.

\para{Inapplicability of the `almost EGS theorem'.}

There is an 'almost Ehlers-Geren-Sachs theorem', which
states that if the observed CMB anisotropy is everywhere
small, then the universe is close to FRW
\cite{EGS} (see also \cite{moreEGS}).
This is contrary to our conclusion above.
The crucial assumption in the proof of the theorem
which is not satisfied in the real universe is that
magnitude of the spatial and time derivatives of
the CMB anisotropy $\Delta T/\av{T}$ is at most
$|\Delta T|/\av{T}$ times the expansion rate\footnote{It
has been earlier pointed
out \cite{Rasanen:2008a} that the theorem is not valid in the
real universe, because it also indicates that the gradient
of the local matter density is at most the density divided
by the Hubble length, times the CMB anisotropy of
(neglecting the dipole) $10^{-5}$.
However, the reason why the theorem fails was not identified.}.
Here $T\propto E$ is the CMB temperature, and the average
on the hypersurface of constant proper time $\av{T}$ can
via statistical homogeneity and isotropy be also understood
as the average over the sky measured at one position.

As we have seen, it is true that $|\Delta E|\ll \av{E}$
at every point, but this does not imply that the derivatives
of $\Delta E$ would be smaller than the derivatives of $E$.
Since the local variation in the expansion rate is of the order
of the average expansion rate, $|\Delta\theta|\sim \av{\theta}$,
it follows from \re{Eeq} that the derivative
$(u^\a+e^\a)\pat_\a(\Delta E)$ is of the order
$|\theta|\av{E}$, rather than $|\theta||\Delta E|$.
Stated the other way round, large derivatives of the
CMB temperature anisotropies do not imply a large variation
in the CMB temperature between nearby points.
The temperature difference is the integral of the derivative
over some length, and as long as the spatial derivative remains
large only over length scales small compared to the inverse derivative,
the difference is small. The small correlation length of spatial
variations is at the heart of the argument which leads from the general
expression for the redshift \re{zt} to the simple relation \re{za}.

\subsection{The distance}

\para{The two-metric.}

Cosmological distances are defined in terms of observations
of light. Two commonly used distance measures are the angular
diameter distance $D_A$, which measures apparent size,
and the luminosity distance $D_L$, which measures
apparent brightness. These distance measures are in
a general spacetime related to each other by the reciprocity
relation $D_L=(1+z)^2 D_A$, so there is only one independent distance
\cite{Ellis:1971}, \cite{Schneider:1992} (page 111),
\cite{Etherington:1933}.
Other distances can be defined by multiplying with
different powers of $1+z$; for discussion of
different distance measures in the context of FRW models,
see \cite{Cattoen, Hogg:1999}.

It is convenient to analyse the null photon geodesics in
terms of the decomposition of the covariant derivative of the
tangent vector, like we did with the timelike dust geodesics.
In the case of timelike geodesics, the tensor $h_{\a\b}$
defined in \re{threemetric}
provides a natural three-metric orthogonal to the flow.
For null geodesics, the situation is more involved,
because it is not possible to construct a metric orthogonal
to $k^\a$ using only $g_{\a\b}$ and $k^\a$:
a new vector field is needed. There is no unique choice,
and different vectors are used in the literature
\cite{Schneider:1992} (page 106), \cite{Sachs:1961, Sasaki:1993}.
For example, we can use the timelike vector $u^\a$,
which is already defined.
It is straightforward to identify the observed area of
a source as the projection onto the two-space
orthogonal to both $k^\a$ and the observer's velocity $u^\a$.
This corresponds to the two-metric
\bea \label{twometric}
  \htt_{\a\b} &\equiv& g_{\a\b} - E^{-2} k_{\a} k_{\b} + E^{-1} u_{\a} k_{\b} + E^{-1} k_{\a} u_{\b} \el
  &=& g_{\a\b} + u_{\a} u_{\b} - e_{\a} e_{\b} \ ,
\eea

\noindent where we have applied the decomposition \re{kdec}
on the second line. The expression
in terms of $e^\a$ is particularly transparent: the two-metric
$\htt_{\a\b}$ spans the subspace of the three-space orthogonal
to $u^\a$ that is also orthogonal to the spatial direction
of the null geodesic $e^\a$.
The two-metric \re{twometric} satisfies
$\htt_{\a\b} u^{\b}=\htt_{\a\b} e^{\b}=0$, $\htt_{\a\b} k^{\b}=0$,
$\htt_{\a}^{\ \c} \htt_{\c\b} = \htt_{\a\b}, \htt^{\a}_{\ \a}=2$.
While conceptually clear, the choice \re{twometric} is not the
most convenient for practical calculations, because $u^\a$ is not
parallel propagated along the null geodesic.
However, the area is independent of the choice of two-metric
\cite{Sachs:1961}. We will mostly keep the two-metric completely
general and use only the condition $\htt_{\a\b} k^{\b}=0$.

\para{The angular diameter distance.}

Given any two-metric orthogonal to $k^\a$, we can decompose
the covariant derivative of $k^\a$ as follows,
\bea \label{photondec}
  \nabla_\b k_\a &=& \thetat_{\a\b} \el
  &=& \frac{1}{2} \htt_{\a\b} \thetat + \sigmat_{\a\b} + k_{(\a} P_{\b)} \ ,
\eea

\noindent where the trace
$\thetat=\htt^{\a}_{\ \b} \nabla_\a k^\b=\nabla_\a k^\a$
is the area expansion rate,
$\sigmat_{\a\b}= \htt_{\a}^{\ \c} \htt_{\b}^{\ \d} \nabla_\c k_\d - \frac{1}{2} \htt_{\a\b} \thetat$
is the shear and $P_\a$ is a vector which depends on the choice
of $\htt_{\a\b}$ and plays no role in what follows. 
We have $\sigmat_{\a\b} k^\b=0$, $P_\a k^\a=0$.
The vorticity is automatically zero (unlike for $u^\a$),
because $k^\a$ is the gradient of a scalar.
Also in contrast to the decomposition of $u^\a$,
the shear is not the symmetric trace-free
part of the full $\nabla_\b k_\a$, but rather the symmetric
trace-free part of $\nabla_\b k_\a$ projected onto the two-space
with the metric $\htt_{\a\b}$. Thus the shear, unlike the area
expansion rate, depends on the choice of two-metric.

Denoting the local scale factor which describes the linear
size of the null geodesic bundle two-space by $s(t,\bx)$,
the area expansion rate is $\thetat=2\patl{s}/s$.
The angular diameter distance is proportional to
the linear size, $D_A\propto s$ (see for example \cite{Sasaki:1993}), so
\bea \label{DA}
  D_A \propto \exp \left( {\frac{1}{2} \int \rmd\l \thetat} \right) \ .
\eea

\noindent As noted, the distance is independent of the choice of the
two-metric $\htt_{\a\b}$ (in particular, it does not depend on
the observer's velocity $u^\a$).

\para{Evolution of the angular diameter distance.}

To determine how the angular diameter distance changes
along the null geodesic, we need the evolution equation of
$\thetat$. As in the case of the redshift, we take a derivative
with respect to $\l$,
\bea \label{Raynull}
  \patl{\thetat} &=& k^\a \nabla_\a \nabla_\b k^\b \el
  &=& k^\a R_{\a\b}^{\ \ \ \b\c} k_\c + k^\a \nabla_\b \nabla_\a k^\b \el
  &=& - R_{\a\b} k^\a k^\b -  \nabla_\b k^\a \nabla_\a k^\b \el
  &=& - 8 \pi\GN \rho E^2  - 2 \sigmat^2 - \frac{1}{2} \thetat^2 \el
  &\equiv& - 2 \mu^2 - \frac{1}{2} \thetat^2 \ ,
\eea

\noindent where $R_{\a\b\c\d}$ is the Riemann tensor, and we have
used the condition $k^\a\nabla_\a k^\b=0$ and the decomposition
\re{photondec}. On the next to last line, we have used the
Einstein equation \re{Einstein}. We have defined
$\sigmat^2\equiv\frac{1}{2}\sigmat_{\a\b} \sigmat^{\a\b}\geq0$
and $\mu^2\equiv4 \pi\GN \rho E^2 + \sigmat^2\geq0$.
Note that $\sigmat^2$, unlike $\sigmat_{\a\b}$, is independent
of the choice of $\htt_{\a\b}$.
The Raychaudhuri equation \re{Raynull} for the null geodesics is
analogous to the Raychaudhuri equation \re{Rayloc} for the timelike
geodesics.

Given the relation \re{DA} and the equation \re{Raynull}, we
obtain the equation satisfied by the angular diameter distance:
\bea \label{DAeq}
  \pat^2_\l D_A &=& - ( 4 \pi\GN \rho E^2 + \sigmat^2 ) D_A \el
  &=& - \mu^2 D_A \ .
\eea

\noindent The right-hand side is non-positive (and the initial
condition for $\pat_\l D_A$ is negative), so the distance is
monotonic along the null geodesic. We also see that the dust
energy density and the photon shear can only make distances smaller
(i.e. objects appear larger and brighter) compared to the non-sheared
vacuum case. In particular, neglecting the null shear would
give an upper bound on the distance.

We need the evolution equation for $\sigmat_{\a\b}$, or at
least for $\sigmat^2$.
The equation for $\sigmat_{\a\b}$ will be different for
different choices of $\htt_{\a\b}$, since the components
$\sigmat_{\a\b}$ depend on $\htt_{\a\b}$. However, the
equation for $\sigmat^2$ can be written in the simple form
\bea \label{sheareq}
  \patl{\sigmat^2} &=& \sigmat^{\a\b} \patl{\sigmat_{\a\b}} \el
  &=& \sigmat^{\a\b} k^\c \nabla_\c \left( \nabla_\a k_\b - \frac{1}{2} \htt_{\a\b} \thetat - k_{(\a} P_{\b)} \right) \el
  &=& \sigmat^{\a\b} k^\c \nabla_\c \nabla_\a k_\b \el
  &=& \sigmat^{\a\b} k^\c R_{\c\a\b\d} k^\d + \sigmat^{\a\b} k^\c \nabla_\a \nabla_\c k_\b \el
  &=& - \sigmat^{\a\b} k^\c k^\d C_{\a\c\b\d} - 2 \thetat \sigmat^2 \ ,
\eea

\noindent where $C_{\a\b\c\d}$ is the Weyl tensor.
We have used the properties $k^\a\nabla_\a k^\b=0$,
$\sigmat_{\a\b} k^\b=0$ and the relations
$\sigmat_\a^{\ \c} \sigmat_{\c\b}=\htt_{\a\b}\sigmat^2$
and $\sigmat^{\a\b}\pat_\l\htt_{\a\b}=0$.
Equation \re{sheareq} is not closed; as in the case of the dust
geodesics, the shear equation of motion cannot be reduced to scalar form.

For practical use, \re{DAeq} should be written in terms
of the observed redshift rather than $\l$, and the contribution
of the null shear term \re{sheareq} should be evaluated.
We will first go through this in the FRW case and then consider
the clumpy situation.

\para{Exact homogeneity and isotropy.}

The symmetry of a FRW universe implies that
(in the appropriate coordinate system)
the diagonal components of $\sigmat^\a_{\ \b}$ should
be equal, and the off-diagonal components should vanish,
since $\sigmat^\a_{\ \b}$ is a spatial tensor.
This implies that $\sigmat^\a_{\ \b}$ vanishes,
since it is traceless.
However, the shape of the source can break the symmetry,
generating non-zero shear.
Since the Weyl tensor vanishes in FRW models, \re{sheareq}
gives $\sigmat^2\propto s^{-4}\propto D_A^{-4}$.
So $\sigmat^2$ can only decrease along the geodesic,
and if it is zero initially, it will remain zero.
We can thus neglect the null shear.

Because of the symmetry, the distance is independent of spatial
position, so $\patl{D_A}=E\pat_t D_A=-E (1+z) H \pat_z D_A$.
Since $E\propto 1+z$ and $\rho\propto a^{-3}\propto (1+z)^3$, 
\re{DAeq} reduces to
\bea \label{DAeqFRW}
  H \pat_z [ (1+z)^2 H \pat_z D_A ] &=& - 4\pi\GN \rho D_A \el
  &=& - 4\pi\GN \rho_0 (1+z)^3 D_A \ .
\eea

\noindent Given the initial conditions $D_A(0)=0, \pat_z D_A(0)=H_0^{-1}$,
the distance $D_A$ is completely determined by the evolution
history $H(z)$ and the present value of the matter density $\rho_0$.
The spatial curvature enters only via it effect on $H$.
For general matter content, $\rho$ on the right-hand side
of \re{DAeqFRW} would be replaced by $\rho+p$, where $p$ is the pressure.

Instead of using the Einstein equation to substitute the
energy-momentum tensor for
$R_{\a\b}$ in \re{Raynull}, we can express $R_{\a\b}$
directly in terms of $H$ and the spatial curvature
$\sR=6 K (1+z)^2$. Essentially, we are swapping $\rho+p$ for
the spatial curvature. This makes it possible to integrate
\re{DAeqFRW} in a closed form, regardless of the matter content
or the theory of gravity: the result only depends on the metric
having the FRW form.
Substituting $4\pi\GN (\rho+p)=-\Hdot+K (1+z)^2$ on the left-hand
side of \re{DAeqFRW} and making the change of variable
$v=\int_0^z \rmd z'/H(z')$, we obtain
\bea
  \pat_v^2 [ (1+z) D_A ]  = - K (1+z) D_A \ ,
\eea

\noindent which integrates into the well-known expression
\bea \label{DAclosed}
  D_A = (1+z)^{-1} \frac{1}{\sqrt{-K}} \sinh\left(\sqrt{-K}\int_0^z \frac{\rmd z'}{H(z')} \right) \ .
\eea

\noindent The fact that the spatial curvature evolves like $(1+z)^2$ in all
FRW universes is the reason why $D_A$ can be written in this
universal form, which depends only on $H$ and $K$.
However, we may equally view $D_A$ as being determined
by $H$ and $\rho+p$. While these quantities are related to the
spatial curvature in a simple manner in the homogeneous and isotropic
case, the situation is different in a clumpy space.

In \cite{Clarkson:2007b}, the universal FRW relation \re{DAclosed}
was formulated as a consistency condition between $D_A$ and $H$.
(A similar consistency condition, but specific to the
\LCDM model, was presented in \cite{Zunckel:2008}.)
If the relation \re{DAclosed} between $D_A$ and $H$
is violated, the metric cannot be the FRW one.
This is a null test: if the condition is not violated,
we cannot conclude that the metric has the FRW form, especially
as recovering $H$ from the observations requires assumptions
about the geometry, with the FRW metric usually adopted to begin with.

\para{Statistical homogeneity and isotropy.}

In a general dust spacetime, the first obstruction to
finding $D_A(z)$ is that no such function exists.
While the angular diameter distance is a monotonic function
of $\lambda$, the redshift is not. As \re{Eeq} shows,
the redshift can both increase and decrease along the null geodesic.
Physically this is clear: in a region which is collapsing
(or strongly negatively sheared in the direction along the null
geodesic), the light gains a blueshift, i.e. a negative redshift.
There is a unique redshift at each point along the null geodesic,
but more than one point may share the same redshift.
Therefore, while the function $D_A(\l)$ exists (and is monotonic),
the expression $D_A(z)$ is not single-valued, and the same redshift
may correspond to several distances. This is true even for a single
geodesic, variation in different directions on the sky aside
(see \cite{Mustapha:1997} for an example).

In a statistically homogeneous and isotropic universe which
expands on average, the redshift accumulated over a
section of the null geodesic passing through a
homogeneity-scale sized region is always positive.
So while $D_A(z)$ does not exist, there is a function
$D_A({\av{z}})$. As before, by statistical homogeneity
and isotropy, $\av{z}$ can be interpreted as either the
spatial average or the average over a section of the
null geodesic longer than the homogeneity scale,
but much shorter than the Hubble scale.

In the case of the redshift, it was possible to directly integrate
equation \re{Eeq} in terms of the variable $\l$, so we could
straightforwardly discuss the smoothing in terms of average
quantities related to the dust geometry.
For the area expansion rate $\thetat$, or equivalently the
angular diameter distance $D_A$, we cannot write down
the solution explicitly\footnote{While
the equation \re{Raynull} for $\thetat$ looks simple, it is a
sub-case of the Riccati equation which does not have a general solution.
Switching to $D_A$ gives \re{DAeq}, the one-dimensional Schr\"{o}dinger
equation, for which there is no general solution either.}.
Nevertheless, we can still follow a similar line of reasoning than
with the redshift, but consider smoothing at the level of the
equation instead of its solution.
The objective is to see whether it follows from statistical
homogeneity and isotropy that the mean of the area expansion
rate $\thetat$ dominates over the variation,
$|\Delta\thetat|\ll|\av\thetat|$ (and if not, what
additional assumptions are needed), and find the
equation relating $\av{\thetat}$ to the average dust geometry.

The source term $\mu^2$ in the equation \re{Raynull} for $\thetat$
has large local variations, just like the right-hand side
of the redshift equation \re{Eeq}. In fact, the variation
is stronger than in the redshift case, since the matter density can
change by orders of magnitude, \mbox{$|\Delta\rho|\gg\av\rho$}.
However, analogously to the redshift case,
$\thetat$ can depend on $\mu^2$ only via the integral
$\int\rmd\l \mu^2$ (and its further integrals).
(This is transparent with the substitution
$\thetat=f(\l)-2\int\rmd\l \mu^2$ into \re{Raynull}.)
We can write this quantity as
\bea \label{muHI}
  \int\rmd\l \mu^2 &=& \int\rmd\l ( 4 \pi\GN \rho E^2 + \sigmat^2 ) \el
  &=& \int_\l\rmd t \frac{1}{E} ( 4 \pi\GN \rho E^2 + \sigmat^2 ) \el
  &\approx& \int_\l\rmd t \frac{1}{\av{E}} \left[ 4 \pi\GN ( \av\rho + \Delta\rho ) \av{E}^2 + \av{\sigmat^2} + \Delta\sigmat^2 \right] \ ,
\eea

\noindent where we have again assumed that vorticity vanishes
so that we can parametrise points along the geodesic uniquely with $t$,
and we have taken into account $|\Delta E|\ll\av{E}$.

For the matter density, we can use the same reasoning as with
the expansion rate in \re{zHI} to argue that the contribution
of $\Delta\rho$ to \re{muHI} is subdominant to the contribution
of $\av\rho$ over sufficiently long distances.
(Via \re{consloc},
the matter density is related to the expansion rate by
$\rho(t,\bx)=\rho(t_0,\bx) e^{-\int_{t_0}^{t}\rmd t \theta}$.)
In fact, the argument is now stronger, because
fluctuations in the density necessarily cancel
due to conservation of mass.
In \cite{Marra:2007b, Marra:2008}, the line average of the
density in a Swiss cheese model was found to be smaller than
the volume average. However, the structures in the model
are not distributed in a statistically homogeneous and isotropic
manner (and they are also unrealistically large).
Randomising the structures leads to a suppression of the
deviation from the FRW result for the distance
\cite{Marra:2008, Marra:2007a, Vanderveld:2008}, and we expect
this to hold also for the density.

There is one additional complication compared to the
redshift case, namely correlation between $\Delta\rho$ and $\Delta E$.
We have neglected $\Delta E$ because it is small
compared to $\av{E}$, and $\Delta\rho$ because its mean
vanishes. However, $\Delta E$ and $\Delta\rho$ are correlated
(the density is anti-correlated with the expansion rate,
which is correlated with the redshift), so the contribution
of the term $8\pi\GN\Delta\rho \Delta E \av{E}$ in \re{muHI}
does not vanish over long distances, and $|\Delta\rho|/\av{\rho}$
can be locally orders of magnitude higher than unity.
However, highly overdense regions can take up only a small
fraction of the volume, because mass is conserved.
In particular, since $|\Delta\rho|/\av{\rho}$
in underdense regions cannot exceed unity, the typical value
in overdense regions is also at most unity, and the overall
mean amplitude of $|\Delta\rho|/\av{\rho}$ cannot compensate
for the suppression factor $|\Delta E|/\av{E}\sim 10^{-5}$.

For the null geodesic shear, there is no such simple argument.
Integrating \re{sheareq}, we have
\bea \label{Weylintegral}
  \sigmat^2(\l) = \sigmat^2(\l_0) e^{-\int_{\l_0}^{\l}\rmd\l' 2\thetat} - e^{-\int_{\l_0}^{\l}\rmd\l' 2\thetat} \int_{\l_0}^{\l}\rmd\l' e^{\int_{\l_0}^{\l'}\rmd\l'' 2\thetat} k^\c k^\d \sigmat^{\a\b} C_{\a\c\b\d} \ .
\eea

\noindent The first term vanishes if the initial shear is zero, and anyway
decreases along the null geodesic, as in the FRW case.
The second term is less straightforward. Using the
decomposition \re{kdec}, the Weyl term in the integrand
can be written as
\bea \label{Weyldec1}
  k^\c k^\d C_{\a\c\b\d} \sigmat^{\a\b} &=& E^2 ( u^\c u^\d + e^\c e^\d + 2 u^{(\c} e^{\d)} ) C_{\a\c\b\d} \sigmat^{\a\b} \el
  &=& 2 E^2 ( E^{\a\b} +  \tilde\epsilon^{\mu(\a} H^{\b)}_{\ \ \mu} ) \sigmat_{\a\b} \ ,\eea

\noindent where we have decomposed the Weyl tensor in
terms of its electric and magnetic components, defined as
$E_{\a\b} \equiv C_{\a\c\b\d} u^\c u^\d, H_{\a\b} \equiv \frac{1}{2} \epsilon_\a^{\ \c\d} C_{\c\d\b\mu} u^\mu$,
and we have denoted the volume element on the two-space
orthogonal to both $u^\a$ and $e^\a$ as
$\tilde\epsilon_{\a\b}\equiv\epsilon_{\a\b\c} e^\c$.
The tensors $E_{\a\b}$ and $H_{\a\b}$ are traceless, symmetric,
and orthogonal to $u^\a$.
Writing the null shear in terms of $e^\a$ and quantities
related to the dust geometry using \re{dustdec} and \re{kdec}
and adopting the two-metric \re{twometric}, we have
\bea \label{Weyldec2}
  k^\c k_\d C_{\a\c\b\d} \sigmat^{\a\b} &=& 2 E^3 ( E^{\a\b} +  \tilde\epsilon^{\mu(\a} H^{\b)}_{\ \ \mu} ) \left( \htt_{\a}^{\ \c} \htt_{\b}^{\ \d} - \frac{1}{2} \htt_{\a\b} \htt^{\c\d} \right) ( \sigma_{\c\d} + \nabla_{(\c} e_{\d)} ) \el
  &=& 2 E^3 E^{\a\b} \left( \htt_{\a}^{\ \c} \htt_{\b}^{\ \d} + \frac{1}{2} e_\a e_\b \htt^{\c\d} \right) ( \sigma_{\c\d} + \nabla_{(\c} e_{\d)} ) \el
  && + 2 E^3 H^{\a\b} \htt_{\a}^{\ \c} \tilde\epsilon_{\b}^{\ \d} ( \sigma_{\c\d} + \nabla_{(\c} e_{\d)} ) \ .
\eea

We could argue that if there is no preferred direction
on the two-space orthogonal to $u^\a $ and $e^\a$,
the diagonal components of $H_{\a\b}$,
as well as $\sigma_{\a\b}+\nabla_{(\a} e_{\b)}$,
in the directions parallel to that space should contribute equally when
integrated over long distances. Then the contribution
of the magnetic Weyl term on the second line of
\re{Weyldec2} to the integral \re{Weylintegral} would vanish.
A similar argument for $E_{\a\b}$ would leave a
product of the off-diagonal terms of $E_{\a\b}$
and $\sigma_{\a\b}+\nabla_{(\a} e_{\b)}$, and it is not
clear why its integral would vanish.
We could try to formulate an argument along the lines
that the contribution of the term involving
$\nabla_{(\a} e_{\b)}$ should vanish in a statistically
homogeneous and isotropic space, since the result
should not depend on $e^\a$ (though this is not obvious),
but this would still leave the dust shear.
There is no clear symmetry reason for $E_{\a\b}$
and the dust shear $\sigma_{\a\b}$, or in terms
of \re{Weyldec1}, the null shear $\sigmat_{\a\b}$,
to be uncorrelated, especially as the Weyl tensor
acts as a source for the null shear.
We cannot get rid of all terms not directly related
to the average dust geometry with symmetry arguments
leading to a lack of correlation over long
distances, unlike in the case of the redshift integral \re{zt}.

However, if the amplitude of the Weyl tensor
is highly suppressed compared to the Ricci tensor (specifically,
to $8\pi\GN\av\rho$), the null shear can be neglected
in \re{muHI}, regardless of the correlations of the Weyl tensor.
In general, the components of the Weyl tensor
are locally not smaller than the matter density.
(In particular, in vacuum regions the Ricci tensor is zero, and the
curvature is manifested entirely via the Weyl tensor;
see \cite{Bolejko:2008a} for an example in a Swiss cheese model.)
However, it is possible that in a statistically homogeneous and
isotropic space the contribution of the Weyl tensor to scalar
observables is small compared to the contribution of the Ricci
tensor, when integrated over scales larger than the homogeneity scale.

From observations, the null shear is known to be small
in the real universe \cite{Munshi:2006}.
The smallness of the shear is theoretically supported
by studies of various models which find only small
effects on the distance when the expansion rate is
the FRW one (see \cite{Rasanen:2008a} for an overview and references).
Nevertheless, since we do not have a general theoretical
argument for the smallness of the Weyl contribution,
we will retain $\sigmat^2$ in the equations.
Even though the Weyl tensor (and its relative contribution to
$\mu^2$) can vary strongly between different regions,
we assume that since the Weyl tensor affects $\thetat$ only via a double
integral, the contribution of the variation $\Delta\sigmat^2$
is small compared to the mean value $\av{\sigmat^2}$,
when integrated over long distances.

In addition to $\int\rmd\l\mu^2$, the area expansion rate
$\thetat$ can explicitly depend on the affine parameter $\l$.
So we should also divide $\l$ into the mean and the variation.
We have
$\int\rmd\l=\int_\l\frac{\rmd t}{E}\approx\int\frac{\rmd t}{\av{E}} -
\int_\l\rmd t \frac{\Delta E}{\av{E}^2}$,
again assuming that vorticity vanishes so that we can parametrise
the null geodesic with $t$, and taking into account
$|\Delta E|\ll\av{E}$. This gives $\av\l$ and $\Delta\l$,
and shows that the variation of $\l$ on hypersurfaces of constant
proper time is small, $|\Delta\l|\ll\av{\l}$.
In the first term, we have dropped the subscript $\l$ to
indicate that it is independent of the specific geodesic.
We can write\footnote{We omit possible dependence
on further integrals of $\mu^2$.} (with some abuse of notation)
\bea
  \thetat\left( \l, \int\rmd\l\mu^2 \right) &\approx& \thetat\left( \av{\l} + \Delta{\l} , \int\frac{\rmd t}{\av{E}}\av{\mu^2} + \int_\l\rmd t \frac{\Delta\mu^2}{\av{E}}\right) \ ,
\eea

\noindent where we have taken into account $|\Delta E|\ll\av{E}$.
The correction terms in both arguments of $\thetat$ are
small compared to the mean, so if we expand $\thetat$ in a Taylor
series around the average values of the arguments, the next order
terms are suppressed and we have the result
$|\Delta\thetat|\ll|\av{\thetat}|$.
(It is crucial that $\thetat$ depends on
$\mu^2$ only via an integral: otherwise the result would not
hold, since $|\Delta\mu^2|\gg\av{\mu^2}$.)

We cannot simply substitute $\thetat=\av\thetat+\Delta\thetat$
into \re{Raynull} and drop all terms involving $\Delta\thetat$,
because, as in the redshift case, we do not in general have
$|\patl{(\Delta\thetat)}|\ll|\patl{\av\thetat}|$. In fact,
$|\patl{(\Delta\thetat)}|\sim |\Delta\mu^2|\gg\av{\mu^2}\sim|\patl{\av\thetat}|$.
However, we can integrate \re{Raynull} once to get an equation
where $\int\rmd\l \thetat^2$ appears instead of $\patl{\thetat}$.
All terms involving $\Delta\thetat$ are then subdominant, and
can be dropped. The resulting equation depends only on the
time $t$, not on the spatial coordinates. Taking a time
derivative, we obtain
\bea \label{Raynullav}
  \av{E} \pat_t\av{\thetat} + \frac{1}{2} \av{\thetat}^2 &=& - 2 \av{\mu^2} \el
  &=& - 8 \pi\GN \av{\rho E^2}  - 2 \av{\sigmat^2} \el
  &\approx& - 8 \pi\GN \av\rho \av{E}^2 - 2 \av{\sigmat^2} \ ,
\eea

\noindent where we have taken into account $|\Delta E|\ll\av{E}$.
This is the equation that we would have gotten by naively
replacing all quantities by their averages in \re{Raynull},
putting $\patl\rightarrow \av{E} \pat_t$, and neglecting
variance and the non-commutation of taking the
derivative and averaging.

We now have all the ingredients for the equation for
the angular diameter distance in terms of the average redshift.
From the relation \re{DA} we get, using similar reasoning as above,
$\av{D_A} \propto \exp \left( {\frac{1}{2} \int \frac{\rmd t}{\av{E}} \av{\thetat}} \right)$
and $|\Delta{D_A}|\ll\av{D_A}$. Inverting, we have
$\av{\thetat}=2 \av{E}\pat_t \av{D_A}/\av{D_A}$.
Recall that $E\propto (1+z)\approx 1+\av{z}=a^{-1}$, and $H=\adot/a$.
From \re{cons} we have $\av{\rho}\propto a^{-3}\propto (1+\av{z})^3$.
Putting the pieces together, we obtain
\bea \label{DAeqav}
  \!\!\!\!\!\!\!\!\!\!\!\!\!\!\!\!\!\!\!\!\!\! H \pat_{\av z} \left[ (1+\av{z})^2 H \pat_{\av z} \av{D_A} \right] &=&  - \left[ 4\pi\GN \av\rho + \av{E}^{-2} \av{\sigmat^2} \right] \av{D_A} \el
  &=&  - \left[ 4\pi\GN \av{\rho_0} (1+\av{z})^3 + \av{E}^{-2} \av{\sigmat^2} \right] \av{D_A} \ ,
\eea

\noindent where
\bea
  \!\!\!\!\!\!\!\!\!\!\!\!\!\!\!\!\!\!\!\!\!\!\!\!\!\!\!\! \av{E}^{-2} \av{\sigmat^2} &=& 2 (1+\av{z})^{-2} \av{D_A}^{-4} \int_0^{\av{z}} \frac{\rmd z'}{H} (1+z') \av{D_A}^4 \times \el
&& \times \left\langle \left[ E^{\a\b} \left( \htt_{\a}^{\ \c} \htt_{\b}^{\ \d} + \frac{1}{2} e_\a e_\b \htt^{\c\d} \right) + H^{\a\b} \htt_{\a}^{\ \c} \tilde\epsilon_{\b}^{\ \d} \right] ( \sigma_{\c\d} + \nabla_{(\c} e_{\d)} ) \right\rangle \ ,
\eea

\noindent where we have assumed that the initial shear is
small and can be neglected, and we have taken into account
$|\Delta D_A|\ll\av{D_A}$.

The equation \re{DAeqav} is our main result for the distance.
It shows that the average angular diameter distance can be
written in terms of the average dust geometry, plus the null shear.
Aside from the null shear term, \re{DAeqav} has the
same form as the corresponding FRW equation \re{DAeqFRW}
in the case when $\rho+p\propto (1+z)^3$, i.e. when the
matter consists of dust and vacuum energy.
When the null shear is negligible, differences between the
distances of a clumpy model and the \LCDM FRW model are
completely encoded in the expansion rate and the redshift.
A clumpy model with the same $H(\av{z})$ (and present-day
matter density) as the \LCDM model
has the same average distance-redshift relation, even though the
spatial curvature will in general evolve quite differently,
as discussed in \sec{sec:av}. (This conclusion disagrees with
\cite{Paranjape:2006b, Larena:2008, Rosenthal:2008},
where the distance was considered using ad hoc
treatments for the spatial curvature; see also \cite{Rasanen:2007}.)
This is in contrast to FRW models with matter other than dust plus
vacuum energy, where the distance deviates from the \LCDM case
because of a different $\rho+p$ in addition to a different $H$.
Of course, this depends on writing the equation for the angular
diameter distance in terms of $H$ and $\rho+p$ instead of $H$
and the spatial curvature. In the FRW case, it is possible to
eliminate $\rho+p$ in favour of the spatial curvature.
Let us see why this does not work in a clumpy universe.

Analogously to the FRW case, we can use \re{Ray} and \re{Ham}
to substitute $4\pi\GN\av\rho=-\Hdot+\av\sR/6+\sQ/2$ on the
left-hand side of \re{DAeqav} and make the change
of variable $v=\int_0^{\av{z}} \rmd z'/H(z')$ to obtain
\bea \label{DAeqavR}
  \pat_v^2 D = - \left( \frac{1}{6} \av\sR + \frac{1}{2} \sQ + \av{E}^{-2} \av{\sigmat^2} \right) (1+\av{z})^{-2} D \ ,
\eea

\noindent where we have denoted $D\equiv (1 + \av{z}) \av{D_A}$.
We could express $\sQ$ in terms of $\av\sR$ using the
integrability condition \re{int}.
However, there is a simple integral \re{DAclosed} in
terms of $v$ only when the right-hand side of \re{DAeqavR}
is independent of $\av{z}$, which requires the expression
inside the parentheses to be proportional to $(1+\av{z})^2$,
such as when $\sQ=0$ and $\av{\sigmat^2}=0$.
(When $\sQ=0$, it follows from \re{int} that $\av\sR\propto a^{-2}$.)

\section{Discussion} \label{sec:disc}

\subsection{Observational and modelling issues}

\para{Spatial curvature and the position of the CMB peaks.}

The position of the CMB acoustic peaks is often
considered to be a measurement of spatial curvature.
The reason is that the peak location in multipole
space corresponds to the apparent size of the
last scattering sound horizon,
which provides a measure of the angular diameter
distance to the last scattering surface at $z\approx1100$
\cite{Durrer:2008} (page 99).
The observed position of the peaks is consistent with spatial
flatness in a \LCDM universe \cite{Komatsu:2008}.

However, the peak position does not imply spatial flatness, even
in a FRW universe. It is clear from the way $H$ and $K$ enter
the distance \re{DAclosed} that for any value of $K$,
it is possible to adjust $H$ to compensate for the spatial
curvature so as to keep the distance fixed.
For example, the peak position is consistent with a FRW universe
with large positive spatial curvature \cite{Spergel:2006}.
Such a model is not viable due to other constraints, such as the
value of the Hubble parameter today. By replacing
the vacuum energy with exotic matter with a time-dependent
equation of state, it is possible to do a similar adjustment
and allow spatial curvature without changing the expansion
history as radically \cite{Ichikawa, Clarkson:2007a}.

In the FRW case, it is true that given the expansion
history $H(z)$, the peak position provides a measurement of
the spatial curvature. The root of this argument is the
relation \re{DAclosed}, which expresses the distance in
terms of the expansion rate and the spatial curvature.
In a clumpy universe the distance is instead completely
fixed by the expansion rate and the matter density
(as well as the null shear), according to \re{DAeqav}\footnote{Note
that we are comparing a clumpy universe
with only dust to a FRW universe with arbitrary matter content.
If we allowed other matter in the clumpy situation,
the distance would also depend on the non-trivial
$\av\rho+\av{p}$. In this case, the Buchert equations would also be
more complicated \cite{Buchert:2001}.}.
As \re{DAeqavR} shows, the expression \re{DAclosed} is
inapplicable due to the non-trivial evolution of the spatial
curvature as well as the fact that
clumping contributes to the expansion rate via $\sQ$.
In terms of the density parameters defined by dividing
the expansion law \re{Ham} by $3 H^2$, the density of matter
and the density of curvature do not sum to unity, because
of the contribution of $\sQ$ \cite{Buchert:1999, Buchert:2003}.
In a FRW model, any additional contribution that changes $H$
enters the distance also via $\rho+p$ (which is related
to the spatial curvature), but that is not the case here.
In the FRW case with arbitrary matter, the spatial
curvature is always proportional to $(1+z)^2$, and the evolution
of $\rho+p$ is not fixed, while in a clumpy dust universe,
$\rho+p$ is always proportional to $(1+z)^3$, but the evolution
of the spatial curvature is complicated.

It is somewhat trivial that the spatial curvature enters the
distance only via its effect on the expansion rate, since the
spatial curvature can be written in terms of $H$, $\Hdot$ and
$\av\rho\propto (1+\av{z})^3$ using the Buchert equations
\re{Ray} and \re{Ham}. However, it is not obvious that
the equation for the average distance \re{DAeqav} depends
only on $H$ and the matter density, or that the dependence on $H$
is the same as in the FRW case, so that we recover the \LCDM
equation (because $\rho+p$ is the same in both cases).

In summary, the CMB peak position can be consistent with large
spatial curvature in a clumpy model. All that is required
is that the expansion history and the matter density today
is sufficiently close to that of the spatially flat \LCDM model.
(Since the Hubble rate enters via an integral,
significant variation in $H(\av{z})$ is still allowed
\cite{trans}.)

\para{The effective equations of state.}

Observations of distances are typically analysed
in terms of an effective equation of state in a FRW model.
(Regarding the dependence of the results on the adopted
parametrisation for the equation of state, see \cite{trans}.)
The equation of state determines the evolution history $H(z)$,
which then gives the distance $D_A(z)$ via \re{DAclosed}.
If we want to express the evolution of a clumpy dust
model this way, there are two different effective
equations of state, because the relation between $H$
and $\av{D_A}$ is different than in FRW models.

For the expansion rate, we can define an effective equation
of state $w_H(z)$ such that a FRW model with this
equation of state would have the same expansion history
as the clumpy model.
Formally this is done by writing  the spatial curvature and
the backreaction variable $\sQ$ as a single component in the
Buchert equations \re{Ray}--\re{cons} \cite{Buchert:2001, morphon}.
For comparison to distance observations, we should introduce
another equation of state $w_D(z)$, defined so
that the resulting $H$ reproduces the real
distance function $\av{D_A}$ when plugged
into the FRW distance formula \re{DAclosed}.
In both cases, we have to make a choice for the 
spatial curvature of the FRW fitting model.
The simplest choice, which does not involve any
loss of generality, is to take the FRW model to be spatially flat.
In general, $w_H\neq w_D$, so the effect of clumpiness
cannot be treated just as an effective source in the FRW
equations.
This limits the usefulness of an effective description
of backreaction in terms of a scalar field \cite{morphon}
(also, $w_H$ can violate the null energy condition, unlike
the equation of state of a scalar field \cite{Rasanen:2008a}).

Until calculations of the impact of structure formation
are accurate to more than an order of magnitude
\cite{Rasanen:2008a, peakrevs}, it is not known
how large the expected difference between $w_H$ and $w_D$ is.
However, even without a theoretical prediction, it is possible
to test the null hypothesis that the equations of state inferred
separately from observations of the expansion rate and distance
do not show any difference. This is the
essence of the FRW consistency check proposed in \cite{Clarkson:2007b}.

At the moment, while distances are measured relatively well,
there are few observations probing the expansion rate as a
function of redshift independent of the distance scale.
The ages of passively evolving galaxies provide an interesting
way to measure the expansion history, but at the moment the
constraints are rather weak \cite{ages}\footnote{In a
space which is not statistically homogeneous
and isotropic, this data does not measure purely the expansion
rate, since the shear also contributes to the redshift
\re{zt}. This was used to constrain local void models
in \cite{Bolejko:2008b}.}.
Another measure is provided by baryon acoustic oscillations,
which are sensitive to a combination of expansion rate and
distance \cite{BAO2}. (This was used to constrain local void models
in \cite{GarciaBellido:2008b}.)
There does appear to be some discrepancy between
the observations of the luminosity distance of type Ia supernovae
and measurements of baryon acoustic oscillations,
but only at the 2$\sigma$ level \cite{More:2008},
so it is not statistically significant.
When analysing the data in the context of FRW models,
this discrepancy would be interpreted as a violation of the
reciprocity relation $D_L=(1+z)^2 D_A$ instead of the FRW
relation \re{DAclosed} between $H$ and $D_A$.

The expansion rate also has a role in the ISW effect, i.e.
deviations of the redshift from the mean value $\av{z}$ in
different directions, as well as in the growth of density
perturbations \cite{lineargrowth}.
The ISW signal is slightly higher than the spatially flat
\LCDM prediction, but the difference is not statistically
significant (it is evaluated as 2$\sigma$ in \cite{Ho:2008}
and 1$\sigma$ in \cite{Giannantonio:2008}).
Neither the ISW signal nor the growth factor can be used
at present to put accurate constraints on $H$ as a function of redshift.

\para{Average observables.}

The equation \re{DAeqav} determines the average angular
diameter distance as a function of the average redshift,
given the average expansion rate. The averages are
taken on the hypersurface of constant proper time.
However, we observe the redshift and the
distance only in one fixed location. 
Nevertheless, for practical purposes, the averages
$\av{D_A}$ and $\av{z}$ do correspond to directly
observable quantities.

In order to model what could be observed in principle,
we would have to know the details of the structures
along each line of sight to calculate the relation
between $D_A$ and $z$ for each direction. As noted
earlier, the distance-redshift relation cannot in general
be expressed as a function $D_A(z)$, because several values
of $D_A$ can correspond to the same redshift.
In practice differences between redshifts
corresponding to the same distance
are likely to be smaller than the observational resolution,
except for nearby sources (or the CMB, for which the redshift
is very accurately measured), because the regions which are
collapsing (or have strong negative shear in the direction
of the null geodesic) are small.
Using the redshift integral \re{zt}, a naive estimate of
the blueshift due to a region one Mpc across which is
collapsing with a rate of the present-day Hubble parameter is $10^{-3}$.
As long as the variations in $D_A$ and $z$ are smaller than the
observational errors, we can safely say that the averages
correspond to the observed quantities.
Observationally, the variation of the CMB peak position 
with direction is known to be small \cite{Hansen:2004}.
Note that standard CMB analysis also predicts only an ensemble
average, and that in practice cosmological observations are
often analysed using an average over the full sky, which by
statistical homogeneity and isotropy corresponds to the average
over the spatial hypersurface.

\para{The Dyer-Roeder approximation.}

An approach where the effect of clumping on light propagation
is modeled assuming that the light rays encounter only a fraction
of the mass in the universe was introduced by Zel'dovich
\cite{Zeldovich:1964} and is known as the Dyer-Roeder
approximation \cite{Dyer}.
In this prescription, the FRW equation \re{DAeqFRW}
for the distance is modified by multiplying the matter density
by a constant $\alpha$, which varies between 0 and 1,
corresponding to a universe where the lines of sight are
completely empty or completely filled with matter, respectively.
The smoothness parameter $\alpha$ was generalised to a function of
redshift in \cite{Linder} to account for the evolution of
structures, and change of the expansion rate due to clumping
was added to the equation in \cite{Mattsson:2007b}.

According to our result \re{DAeqav}, clumping is irrelevant
for the contribution of the matter density, which
is always proportional to $(1+\av{z})^3$,
with no extra prefactor. The reason is that mass is
conserved, so if the line of sight goes through an underdense
region somewhere, it must correspondingly go through overdensities
elsewhere when considering distances of the homogeneity scale
or larger, as discussed in connection with equations
\re{zHI} and \re{muHI}.
The Dyer-Roeder parameter $\alpha$ is always unity, and clumping
enters instead by changing the expansion rate.
In addition, there is the null shear term; if interpreted as
an effective, redshift-dependent $\alpha$, it would correspond
to $\alpha>1$, contrary to the Dyer-Roeder case.
In summary, the Dyer-Roeder prescription does not correctly
describe the effect of clumping.

One possible caveat is that we have not taken into account
the possibility that the observer
or the sources could be in a special location, or that
observations could be made along special lines of sight.
For example, we would expect supernovae to be preferentially
located in very overdense and thus highly untypical regions.
The possibility that observations might be biased towards
empty lines of sight has been brought up in \cite{Mattsson:2007b, Weinberg:1976}.

The effect of the location of the observer is likely to be small,
because the deviation from the mean is significant
only over regions which are small compared to the overall
distance travelled by the light, and the amplitude of the
deviation is typically not correspondingly large (except in very
special locations, such as near a black hole).
This is another way of saying that the homogeneity scale is small.
For the same reason, the location of the sources is not expected to
have a large impact, though there could be a secular effect,
as the degree to which the source locations are untypical
could evolve with redshift.
In any case, cosmological observations rely on various
different sources, not only supernovae.
Since different observations roughly agree, these kind
of selection effects must be subdominant.
This is also an argument against large effects due to
special lines of sight used in observations. In particular,
the CMB covers the full sky, so it is not subject to this kind of
bias (apart from some uncertainty in the direction of the Galactic plane).
Furthermore, since most cosmological observations (apart from nearby
objects) are made over scales much larger than the homogeneity scale,
the variation in the density along lines of sight should be small,
and empty lines of sight should be very rare. Note that a clear
line of sight is not necessarily empty, because most of the dust
is dark matter, which is transparent.
(See \cite{More:2008} and references therein for more on cosmic
transparency.)

\subsection{Conclusion}

\para{Summary.}

It was conjectured in \cite{Rasanen:2008a} that light
propagation in a statistically homogeneous and isotropic
dust universe which may contain non-linear structures
can be expressed in terms of average geometrical quantities
(namely the expansion rate and the spatial curvature),
assuming that the observer is not in a special
location, and that structures have realistically small sizes.
As reviewed in \cite{Rasanen:2008a}, the literature on light
propagation is mostly in agreement with this statement, but
there had been no proof thus far.

We have now derived the equation for the angular diameter
distance, assuming that the dust universe is
statistically homogeneous and isotropic as well as rotationless,
and that structures are small and their distribution evolves
slowly compared to the time it takes for light to cross the
homogeneity scale.
It follows from these assumptions that the distance can be
expressed in terms of the average dust geometry, apart from
a term related to the null geodesic shear. 
Of the average geometry, the only term that is required
is the average expansion rate, along with the present
value of the matter density.
In particular, the spatial curvature enters only via its effect
on the expansion rate, unlike in Friedmann-Robertson-Walker
(FRW) models with arbitrary matter content.
Therefore, significant spatial curvature is not necessarily
inconsistent with the position of the cosmic microwave background
acoustic peaks.
If the expansion history and present-day matter density
of a clumpy model is close to that of the spatially flat
\LCDM FRW model, the peak position will also be near the
\LCDM case, and thus consistent with observations.
This is important, because accelerated expansion due to
structure formation involves large negative spatial
curvature \cite{Rasanen:2005, Rasanen:2006b, Rasanen:2008a}.
The result also shows that the clumping is not correctly
described by the Dyer-Roeder approximation,
which changes the evolution of the matter density
(which is in fact fixed) and misses the change of
the expansion rate.

To complete the proof that the redshift and the distance
can be expressed in terms of the average geometry alone,
it would be necessary to show that the contribution of the
null shear can be neglected. Observationally, the shear is
known to be small \cite{Munshi:2006}, but this should be shown
to follow from statistical homogeneity and isotropy (or, if this
is not the case, the required additional assumptions should be identified).

Since the distance-redshift relation is determined by the
average expansion rate, we should calculate the
average expansion rate in a realistic model
to evaluate the effect of structure formation on
cosmological observations.
Using the Buchert equations \cite{Buchert:1999}, this
backreaction can be determined from purely statistical quantities
(the variance of the expansion rate and the average dust shear).
Evaluating these quantities in a realistic model is a
challenging task, especially as the backreaction is a
general relativistic effect related to spatial curvature
and has no counterpart in Newtonian gravity
\cite{Buchert:1995, Buchert:1999, Buchert:2007, Rasanen:2008a}\footnote{Regarding
the differences between Newtonian gravity and the weak field limit
of general relativity, see \cite{Ellis:1971, Ellis, Senovilla:1997, Szekeres}.}.
A first step was taken in \cite{Rasanen:2008a, peakrevs},
where the average expansion rate was calculated in a
semi-realistic model with an evolving distribution of structures.
To compare with observations of distance and
the expansion rate in detail, a more rigorous treatment
is needed, with well-quantified errors.
After that, the calculation of fluctuations around
the average should follow, in order to evaluate the Integrated
Sachs-Wolfe effect and the growth of density perturbations.

Before detailed analytical results on the effect of structure
formation are worked out, it is possible to make general tests.
The relation between the expansion rate and the distance scale in
clumpy models is different than in FRW universes.
Therefore, observations which probe the expansion rate and the
distance separately can be used to test the null hypothesis that
the two are related by the FRW consistency condition, as proposed
in \cite{Clarkson:2007b}.
Comparing observations of type Ia supernovae, baryon acoustic
oscillations and the ages of passively evolving galaxies
seems promising in this respect.

\ack

I thank Thomas Buchert for discussions and comments
on the manuscript and Ruth Durrer, Teppo Mattsson and
Subir Sarkar for helpful discussions.  \\

\setcounter{secnumdepth}{0}

\end{document}